# Automating the Horae: Boundary-work in the age of computers

Luis Ignacio Reyes-Galindo[1]
*School of Social Sciences, Cardiff University, Wales.*

**Abstract**

This paper describes the intense software filtering that has allowed the arXiv e-print repository to sort and process large numbers of submissions with minimal human intervention, making it one of the most important and influential cases of open access repositories to date. The paper narrates arXiv's transformation, using sophisticated sorting/filtering algorithms to decrease human workload, from a small mailing list used by a few hundred researchers to a site that processes thousands of papers per month. However there are significant negative consequences for authors who have been filtered out of arXiv's main categories. There is thus a continued need to check and balance arXiv's boundaries, based in the essential tension between stability and innovation.

**Keywords**
*boundary work, open access, arXiv, physics*

> Arete (Excellence) with garland-loving Eukleia (Good Repute) steers the city, she and wise Eunomia (Good Order in civic government), who has festivities as her portion and guards in peace the cities of pious men.
> Bacchylides, Fragment 13[2] (5th century B.C.)

> Edges, Borders, Boundaries, Brinks and Limits have appeared like a team of trolls on their separate horizons. Short creatures with long shadows, patrolling the Blurry End.
> A. Roy, The God of Small Things

### ❧Introduction: Physics and the ubiquity of non-peer reviewed communications❧

Although the importance of preprints in communicating physics and mathematics is well-known to historians of science (Bohlin, 2004; Bourne and Hahn, 2003; Kaiser, 2009, 2011; Lewenstein, 1995; Wykle, 2014), there is little STS literature on preprint usage and distribution. This paper adds to the study of preprint cultures a micro-sociology of the site *arXiv*, the dominant 'e-print' repository and distribution site for the hard sciences, including most of the sub-fields of physics. Approximately 70-80% of physics journal papers are first made available on arXiv, the site has accumulated more than one million preprints, and arXiv literature is now commonly cited in peer-reviewed journals even though the arXiv versions may not have passed through a peer-review process (Larivière et al., 2014). In some areas, like high-energy physics (HEP), arXiv is by far the leading source of technical literature, and most citations in published papers are arXiv preprints (Cronin, 2003). arXiv is an intrinsic part of the physics publishing world, despite peer-reviewed journals' continuing to provide the highest-grade imprimatur (Rieger and Warner, 2012).[3] Although a number of bibliometric studies have analysed the positive effects of repositories on academic communication (Björk et al. 2014; Gargouri et al., 2012; Laakso, 2014; Larivière et al., 2014), empirical, qualitative investigations of so-called 'green' open access, such as the present work, so far do not exist.

---

[1] LuisReyes@ciencias.unam.mx

[2]Trans. Campbell, Vol. Greek Lyric IV. I thank one of the paper's anonymous reviewers for correctly pointing out that of the three deities mentioned, Eunomia is the only true *Hora*, a keeper of the gates of heaven, along with Dike (Justice, Fair Judgements and the rights established by custom and law) and Eirene (Peace). Eukleia and Arete are *daimona*, or minor goddesses, but the title and headings were kept as is mostly for stylistic purposes.

I begin with an historical introduction to arXiv founder Paul Ginsparg's implementation of software-led thematic sorting of submissions, and go on to discuss the human-led filtering by 'subject moderators' that occurs after automatized filtering. I then describe how, in conjunction with an automated analysis of style, the sorting is repositioned by arXiv as a filter for submission 'quality'. In the conclusion I discuss the consequences of being 'filtered out' of arXiv's main subject categories and the stigma attached to this process. I chose Ginsparg as a key informant, given that from 1991 until 1994 – arXiv's formative period – he was the sole developer of the site's software backbone. His approach of using DIY, maximally automated, computerized content-filtering to match as closely as possible the disciplinary structure of physics continues to guide arXiv to this day. Though arXiv is now managed by Cornell Library and directed by a Scientific Board, arXiv remains an evolved but direct descendant of Ginsparg's original design.[4]

I identified arXiv physics moderators through the website's public information page and contacted them directly. Over a two-week period in 2015, I interviewed five moderators, who were chosen for geographical convenience. The work of these moderators represents a range of areas in physics; not explicitly being listed anywhere on the site these were therefore not a factor in the selection process. Two senior editors of high profile scientific journals provided feedback on my analysis of the relationship between editorial practices in mainstream physics and arXiv editorial practices. To research the issue of arXiv's boundary making by including views of those 'filtered out' of the site, I interviewed and corresponded with key informants outside the borders of mainstream physics. Interviewees included: independent scientist and creator of the alternative e-print server 'viXra' Phil Gibbs, a well-known senior physicist working on unorthodox topics, two junior scientists who had troubles posting on arXiv, and mainstream scientists who had experience collaborating with fringe scientists. I also corresponded at length with fringe scientists. This research is an overlapping part of two ongoing research projects, which look at the wider ecology of physics publishing, open access and the ecology of fringe science. The projects are informed by ethnographic fieldwork and interviews with 'independent scientists' that will be included in future publications.

---

[3] Bohlin (2004) has outlined the arXiv transition from paper-based to a purely electronic form. Gunnarsdóttir (2005) has carried out an ethnomethodology of arXiv administrator's daily practice. Collins (2014) and Collins et al. (manuscript) have outlined the role of arXiv within the wider ecology of physics in and far outside mainstream science.

[4] P. Ginsparg, 'arXiv at 20: The price for success'. Conference video, Cornell University Library. Posted on 5 December 2011. Available at: http://www.cornell.edu/video/arxiv-at-20-the-price-for-success (accessed 3 June, 2015).

## ❧Eunomia: Preprints, librarians and arXiv❧

Since its early days, the high-energy physics community has exhibited sociological characteristics that have made preprint exchange a critical communications tool. HEP began as a relatively small community of a thousand researchers, with a socially tight-knit yet geographically dispersed membership. HEP physicists have historically exhibited high levels of computer literacy. Finally, the division of labour between HEP experimenters and theoreticians has required speed in the communication of novel results that exceeds the means of traditional printed journals (Bourne and Hanh, 2003; Kaiser, 2009; Gentil-Beccot, 2010; O'Connell, 2000).

The story of arXiv begins in 1989, with an email list that distributed full TeX articles to a core group of 150-180 HEP physicists with interests in two-dimensional gravity and conformal field theory.[5] TeX is a highly portable, device-independent computer typesetting language that produces high quality mathematical papers encoded in simple text files, widely used in physics and mathematics (Beebe, 20014). Although TeX files are small in size by today's standards, at the time they were large enough to fill up expensive disk space quickly. To deal with these disk limitations, Ginsparg wrote a software script that plucked the basic bibliographic information from each submission to the list and assigned each paper a unique identifier number based on the submission date. Instead of sending out the entire files for the full articles, a daily digest announcing the papers submitted during that day, along with instructions for retrieving the full text via automated email request, was automatically compiled and emailed. List members, if interested in reading a particular preprint in full, could retrieve the full TeX paper from a remote server, without using up disk quotas for unwanted papers. Although originally meant temporarily to store only the latest preprints, by the early 1990s, disk space had become sufficiently affordable that keeping a large amount of preprints in a central server was feasible, and Ginsparg was persuaded to maintain a permanent preprint repository (Ginsparg, 1994). In 1991, the system was moved to a dedicated machine at Ginsparg's institution, the Los Alamos National Laboratory (LANL), where Ginsparg requested a top-level email address at *lanl.gov* but was instead given a dedicated machine and the *xxx.lanl.gov* hostname, the original email address under which arXiv ran.

The relevance of the system can be seen when we consider the disorder of HEP preprint sharing before arXiv entered the picture. According to Rosenfeld et al.'s (1970) survey of HEP preprint-sharing infrastructures, approximately 3,000 new preprints were being circulated each year at that time, but with 'little attempt at any uniform distribution policy', despite organized efforts from the major HEP laboratories in Europe and America to improve sharing and indexing. Ginsparg's distribution scheme, however, also serendipitously provided the HEP community with a much needed system to store, classify and cite preprints in a stable, orderly and enduring form (Ginsparg, 1994) and brought order to what had been a ubiquitous but anarchic preprint exchange culture, in which preprints were used often, but were hard to locate, store, classify, and therefore cite (Gentil-Beccot, 2010; Wykle, 2014). As one member of arXiv's moderation team explained:

> Moderator 1: arXiv is not a regular publication path. It's more like a bookshelf. You try to sort things into the shelf properly. And that's good for both those people who publish the book, because the book can be found, but also for the readers because they will find the books that they care about.

---

[5] See J. Cohn. 'Some arXiv preprint history'. Available at: http://astro.berkeley.edu/~jcohn/arxiv_hist.html (accessed 8 March, 2016).

The eventual inclusion of Ginsparg's *hep-th* preprint identifiers in the SLAC-SPIRES database, which had been curating preprint information since 1969 and remains a widely used bibliographic resource, allowed the site to become visible to a much wider audience beyond HEP. SPIRES also importantly became the stepping-stone that gave arXiv early access to what would become a revolutionary technological resource, the World Wide Web. As it did the entire Internet in the early 1990s, the Web transformed arXiv from a local resource with access limited to academics and government institutions into a publicly accessible object (Abbate, 1990). The email-centred distribution list became *http://xxx.lanl.gov*, a site that could be read by a globally distributed audience, both academic and non-academic. As an outcome of the Web transition and over a quarter-century of constant growth, today's renamed site, *http://www.arXiv.org* handles an astounding 10 million downloads per month and registers 580 million historical preprint downloads.[6]

### ❧Eukleia: The friends of our friends❧

Despite its exposure to the wider public of the Internet, arXiv still aims to serve the needs of *scientific* communities. arXiv defines itself, in Ginsparg's original spirit, as 'an openly accessible, moderated repository for scholarly articles in specific scientific disciplines' so that 'material submitted to arXiv is expected to be of interest, relevance, and value to those disciplines'. arXiv is 'distinct from the web as a whole, because arXiv contains exclusively scientific content' and 'is not a repository for otherwise unpublishable materials'.[7] arXiv management promotes a model of science solidly based on a notion of community-produced knowledge. arXiv administrative white papers, for example, describe the site as a 'sociotechnical system', citing STS literature (Rieger and Warner 2010). As Ginsparg explained regarding authors whose papers are meant to populate and regulate arXiv:

> Ginsparg: My criterion, based on training and background and publication record, is that we regard as the community the people we know and respect and the people they refer to us. You know, our friends and the friends of our friends in this generalized sense and that defines the community.

Critically, arXiv responds to the very diverse needs of a spectrum of esoteric expert communities in the wider disciplines of physics, mathematics, computer sciences and other quantitative 'hard' sciences, broken up into highly specific subject categories. The branching off into different subject areas linked to small-sized communities, in the order of 1000-2000 core members, according to analyses of arXiv access logs, is historically embedded in the arXiv system. As arXiv became better known to scientific communities beyond HEP that also wanted to use the system, new subject areas were included and 'subject moderators' were brought into the publication workflow, to regulate postings based on their own communities' interests:

> Ginsparg: [Subject moderators were there] pretty much from the outset. Everyone who suggested a new subject area (the first was Dave Morrison suggesting *alg-geom* [category] in '92, which became *math.AG* later in the 90s) became its moderator. Later larger expansions (e.g. [computer science]), partitioned themselves into subject areas with pre-assigned moderators for each.[8]

---

[6] 'arXiv monthly download rates'. Available at: http://arxiv.org/stats/monthly_downloads (accessed 2 April, 2015).

[7] 'arXiv primer'. http://arxiv.org/help/primer; 'Moderation', arXiv help section. http://arxiv.org/help/moderation; 'The arXiv endorsement system', arXiv help. http://arxiv.org/help/endorsement (all accessed 8 March, 2016). See also comments by Ginsparg in Collins et al. (2015).

To date, arXiv names 135 moderators on its public site, but lists only the wider disciplinary fields, not the specific subject categories, to which moderators are assigned. In their roles as the 'embodiments' of a community's specific research interests, moderators are assigned to carry out two broad regulatory roles. First, they ensure that submissions meet a minimal standard of interestingness and reject submissions that do not meet the referee-abilty criterion: if a paper flagged for moderation would not even be sent to referees by a competent journal editor in that subject area, moderators should reject it. Second, they may decide that the subject category proposed by a potential poster is unsuitable and reclassify the submission into another category, without rejecting it. Moderators can limit 'cross-listing' rights, that is, the ability to announce a submission within other subject areas in arXiv.[9] If in doubt, mods are explicitly asked to err on the side of acceptance rather than on the side of rejection. As one moderator explained:

> Moderator 2: With rejections we are careful. It is actually quite rare that papers are rejected …. If something is not an obvious case for being rejected, I accept it …. I think I've rejected about three papers in the ten years that I've been doing this. This is extremely rare. Maybe five … but that's the order of magnitude, whereas re-categorization happens rather frequently …. I do understand if you want to publish something in *hep-ex* and the *hep-ex* moderator tells you, 'this doesn't go in here', that is problematic, but I'm happy I'm not in this position![10]

Moderators reported relying on simple clues to grade 'interestingness' to a community, such as recognising the names of peers in their specialty field, seeing titles related to standard contemporary research topics, use of field-specific methods, appropriate referencing, etc. A small percentage of papers require a glance at the abstract, and only a very few are read in full. However, as has been described in analyses of the practices of journal editors-in-chief in physics (Reyes-Galindo, 2011), moderators recounted that they had developed an intuitive 'feel' for distinguishing papers that are acceptable from those that are blatantly unacceptable, a feel that goes beyond the simpler, immediate clues:

> Moderator 3: I would say in this arXiv work I use – even unconsciously – a lot of my experience as a journal editor that is much, much longer, because I've been doing this for ten years … or more than ten years. I think I have developed some kind of … what can I say? A 'neural network' that identifies quite quickly bad papers … what is patently bad. Those usually I do quickly. Papers that are in between … it depends. They may take more time. Some may take only a few minutes to quickly go through. Because the level of filtering of arXiv is not a proper review in scientific terms, it's not the time to go in-depth into a paper, but in some cases, I have spent a lot of time – and by a lot I mean two or three hours – to really understand some cases where I was not evidently convinced that this was a scholarly paper.

---

[8] Personal communication, 24 March 2015.

[9] P. Ginsparg, personal communication, 12 February, 2015.

[10] This is consistent with an exhaustive analysis of arXiv moderation decisions on the entirety of physics submissions from July 2014- June 2015, according to which just slightly less than 1% of papers are outright rejected (Collins et al., manuscript). The figure may be significantly higher for some of the more publicly visible categories, such as *hep-th*.

Moderators reported that they would accept papers from familiar authors with whose conclusions they might likely disagree, so long as the authors were immersed in debates relevant to the HEP research community. Following their only available guidelines and FAQs, moderators were unanimous that, in cases of doubt over whether a particular paper should be rejected or not, it is always preferable to err on the side of acceptance.

Despite its moderators' being explicitly told to err on the side of acceptance, and its relatively low rejection rate, the site has been the subject of harsh criticism from a number of individuals. Most notably, Nobel Prize laureate Brian Josephson has censured arXiv with accusations of 'blacklisting'. Allegedly, if even accepted, submissions on fringe physics topics (e.g., cold fusion, consciousness and the paranormal) are automatically relegated to the less prestigious 'general physics' category. The website *arXiv Freedom* also lists cases and names of independent scientists who accuse arXiv of 'blacklisting', 'repression', 'victimization' and 'abuse'. When interviewed, Ginsparg vehemently rejected claims of 'blacklisting' (understood as the targeting of specific authors), although he admitted that the system immediately flags articles that deal with topics known to have long 'gone past their sell-by date', an editorial practice that is standard in the physics publishing world (Collins et al., manuscript). One of the formative intentions of science 'beyond the fringe' is that it is *primarily oppositional* (Collins et al., manuscript). Because arXiv deals with submissions tailored for the specific research communities represented by its subject categories, it would be extremely hard for 'oppositional physics' to make its way into arXiv at all. The exclusion of oppositional physics from arXiv is therefore not a subject of the present paper. Other alternative outlets exist, such as the much smaller viXra site, but are known amongst professional physicists to contain material of no interest.[11]

**❧The filters: Automating the Horae❧**

A remarkable feature of the arXiv system is its efficiency in terms of number of submissions processed vs. personnel employed. 100,000 submissions per year are managed by a small core technical team of fewer than six full-time employees, along with the input of the volunteer subject moderators.[12] Given that the rate of submissions has effectively grown several orders of magnitude faster than the number of technical personnel, arXiv's operation would not be viable within its current operating budget without heavy automatization and computer-aided filtering.[13] Thus, in its public information site, arXiv describes itself as a 'highly-automated electronic archive and distribution server for research articles'.

Moderator time is a particularly precious commodity, given that 'mods' are all unpaid volunteers.[14] To ensure minimal waste of moderator time, before papers reach moderators they must go through a series of barriers that include a complex, home-grown and constantly evolving software filtering system that can weed out papers that community-oriented objectives deem blatantly uninteresting (i.e., papers not identifiable as interesting to at least one of the specific subject communities). Given that arXiv works using so-called 'negative moderation' – a paper that is not flagged will be published automatically with or without moderator approval – it is critical to strike a filtering balance between minimizing moderator intervention (catching the most outliers) and filtering out legitimate papers because of idiosyncrasies. Likewise, filters that are too lax can increase moderator workload, or lead to a flood of uninteresting papers. As with other examples of modern, electronic information dissemination systems, arXiv faces a critical tension between leaving out too much and admitting too little (Hanseth et al., 1996).

---

[11] See Kelk D and Devine D (2012) for a limited bibliometric comparison between arXiv *viXra*.

[12] The figure of 6 full-time equivalents is taken from the arXiv budget. See http://arxiv.org/help/support. This figure includes user support, programming, systems maintenance, management, moderator coordination and administration, so the technical staff is in reality less.

[13] See 'arXiv monthly submission rates' http://arxiv.org/stats/monthly_submissions for historical data on monthly submission rates. Total monthly submissions have increased almost linearly throughout arXiv's history, by about 40,000 every year.

[14] Except for the gen-ph category moderator.

At the first stage of the process, a paper is analysed in terms of technical points, sometimes making it necessary for the administration team to double-check the submission. If it is not flagged on such technical grounds, the paper then passes through two much more complex filters. The system is based on machine-learning software that is constantly being 'trained' and tinkered with, and there are no firm, enumerable criteria for flagging a paper that can be set out in detail (Collins, Ginsparg and Reyes-Galindo, manuscript). Ginsparg explained:

> Ginsparg: Flagging could be for anything. It could be for somebody who systematically submits to the wrong [category], or systematically reverses first name and last name so it has to be corrected, or systematically submits two *TeX* files so there's a double copy of the text. Anything imaginable that just requires a double check. When this was implemented it was completely benign because this thing would just put up a flag for any of these various things.[15]

If they are 'flagged' by the system on technical grounds, submissions must enter a moderation queue for further technical scrutiny, before either proceeding to the next filtering stages or possibly being rejected. With the technical filter, arXiv manages to implement, in relatively simple steps, submission barriers that weed out most submissions from outside the world of institutionalized physics (Gunnarsdóttir, 2005, 2010). At the same, arXiv allows outsiders with minimal social links and technical connoisseurship of mainstream physics to be admitted through 'endorsement' by certified users (Collins et al., manuscript).[16] It is after the initial technical filtering that a sociologically intriguing, fully automated, linguistic and semantic filtering process is implemented in two stages.[17]

---

[15] As Collins et al. (manuscript) have summarized, the initial barriers faced by aspiring posters to the higher-end categories are based on criteria of basic technical skills (ability to write TeX code proficiently), interpersonal trust (submitters must be endorsed by two active submitters to the higher-end categories) and institutional certification (submissions from individuals based at certified academic institutions are given considerably less scrutiny, or none at all, than non affiliated ones) that mirror long standing practices in the physics publishing ecology.

[16] Endorsement only applies to specific subject categories and endorsers can only endorse for the categories they are allowed to post in. Endorsement can also be taken away if one is not an active poster to a category one is endorsed in. See 'General information about arXiv', available at: http://arxiv.org/help/general (accessed 30 June 2015).

[17] For an ambitious proposal for an ecological sociology of algorithms and how these shape and are shaped by large-scale social structures, institutions and practices, see MacKenzie (2014).

*a) The 'subject classifier'*

The first text filter is related to the distribution of 'technical' keywords in a text, which are used to sort papers by subject area (through what is known in computer science as a 'naïve Bayesian filter'):

> Ginsparg: [First] you just use the technical words, say, 'superconductivity', anything that isn't in that list of top few hundred most frequent words .... Not surprisingly, if you look at those technical words you can build a very good subject-area classifier, because now you're just throwing out stylistic differences. You're just focusing on content. You see whether the article contains 'black hole', or whether it contains 'superconductivity', or whether it contains 'databases'. You can get very, very accurate, fine-tuned classification.

Moderators were generally satisfied by the results of the subject classifier, and moderators who had been active for a significant number of years commented that the workload had visibly decreased after subject classifiers were implemented. Mods receive the numerical results of the classifier and, particularly for the more wide-ranging categories, find the subject classifier a useful tool:

> Moderator 1: [The subject classifier] is a huge benefit because then I don't need to look at the [target reclassification] categories myself. It tells me what those are. I don't always agree with this thing. I don't always react on its suggestions, but if I think that something should be moved somewhere then usually the system tells me right away which of the categories I'm thinking about.

> Moderator 4: [T]he filter put in front of me is very effective. I get essentially no crap on my table. That is definitely a change that I appreciate, which means that I can definitely focus on the topic and on the classification of the topic. I'm very glad of that, because I don't want to put too much time in this business.

As a tool for making reclassification decisions, arXiv functions for moderators as a system to support quick sorting calls; a 'decision support system' (King and Star, 1990) or an 'information infrastructure' (Bowker and Star, 1999). And yet, although originally intended only as an instrument for minimizing mod workload by correctly matching submissions to disciplinary categories, the subject classifier has had an unforeseen spinoff: it can spot fringe contributions by marking papers that do not belong to any specific category. These outliers are flagged because they do not fit the keyword pattern of *any* specific discipline. Ginsparg explained:

> Ginsparg: The first thing I noticed was that every once in a while the classifier would spit something out as 'I don't know what category this is' and you'd look at it and it would be what we're calling this fringe stuff. That quite surprised me. How can this classifier that was tuned to figure out category be seemingly detecting *quality*? [emphasis mine]

Or as one moderator explained concerning one of the very few submissions he recalled rejecting:

> Moderator 1: One of the papers we recently rejected was because it didn't fit any of the categories. That was hinting at a possible problem with the submission because it had very little scientific content. It was basically a description of what you can do given a certain problem and that's not necessarily interesting for any of the categories in arXiv. That's a useful discriminator actually. If you don't know exactly where to put it maybe it's because there is really nothing there to put.

The underlying idea in both quotes is that 'quality' work occurs within the specific expert communities that make up arXiv. The equivocation of positing 'interestingness' as a proxy for 'quality' is what Gieryn (1999) has defined as *expulsion boundary work*. Through these strategies, 'all sides seek to legitimate their claims about natural reality as scientifically made and vetted inside the authoritative cultural space', while simultaneously the dominant cultural space 'displays its ability to maintain monopoly over preferred norms of conduct' (Gieryn, 1999: 16).

*b) The 'stylistic' filter*

The second feature of the software filtering relies on the usage of 'stop words' in a text, which are the most common used words in a language. Stop word frequencies in texts approximate closely the power law distribution known as Zipf's Law, a basic result in linguistics and information retrieval sciences. Ginsparg uses stop word frequencies further to spot 'outliers':

> Ginsparg: The first order effect is there are two significant contributors to variations [from Zipf's Law]. One is the subject area one is writing about. The second is the native language of the speaker …. But [outliers] also show up in the stop word distribution, even if the stop words are just catching the style and not the content! They're writing in a style which is deviating, in a way. I find this by doing a principal component analysis and dimensional reduction. You do this and you find, just by using the stop words, here are the physics articles, here are the computer science articles. You can see differences within, and then you find these outliers.[18]

Like the quoted moderator's 'neural network', this filter tunes into stylistic patterns that, independent of actual content, define the 'correct' writing style of subfields. In conjunction with the semantic filter, this leads to a sophisticated analysis that includes content *and* style:

> Ginsparg: What it's saying is that people who go through a certain training and who read these articles and who write these articles learn to write in a very specific language. This language, this mode of writing and the frequency with which they use terms and in conjunctions and all of the rest is very characteristic to people who have a certain training. The people from outside that community are just not emulating that. They don't come from the same training and so this thing shows up in ways you wouldn't necessarily guess. They're combining two willy-nilly subjects from different fields and so that gets spit out.

Though the technical filters are a sufficient barrier to impede most amateur and non-professional authors to upload papers, the automated filters place an even greater demand on aspiring authors. They must *also* have significant amounts of tacit knowledge acquired through enculturation in order to be given the 'free pass', without flags, to avoid close moderator scrutiny. By also policing the boundary of aesthetics, the stylistic filters further align the locus of legitimate interpretation with the group that produces the knowledge, without requiring that group to engage with the technical dimensions of knowledge. The appeal to an intangible 'style' that outsiders cannot emulate without being immersed in a community – the appeal to aesthetic performance of 'proper' physics – recalls Bourdieu's (1979) differentiation between high and low culture and the discourse of 'pure' and 'barbaric' aesthetics through which dominant cultural groups define and control aesthetic domains.

---

[18] The indexing of texts in terms of 'latent' semantic differences is a well-established field in applied linguistics (Deerwester et al., 1990; Hofmann, 1999; Landauer and Dumais, 1997). Computer-based natural language processing has been used with positive results to carry out stylistic analyses such as authorship attribution (Juola, 2006); further work is planned by the author on the possible links between these stylistic differences and Collins' (2010) concept of collective tacit knowledge.

*c) Reclassificaition*

Moderators, when faced with a submission that has been flagged, have essentially four options: 'publish into original category', tag for reclassification and 'publish into another high-profile category', 'reject' or tag as 'gen-ph (or math.GM) at best'. If tagged for reclassification (as opposed to being rejected outright), a paper has two possible destinies. One is to be reclassified into another high-status category that is of relevance to another core community. The other is to be reclassified into one of two categories, general physics (*gen-ph*) or general mathematics (math.GM). These arXiv categories unofficially serve as 'quarantine' areas for irrelevant or low-quality submissions that, although from legitimate posters (i.e. people who have accredited scientific affiliations or meet endorsement requirements), are judged to be uninteresting to all communities.[19]

The division of arXiv into *gen-ph* (and math.GM) and all other 'legitimate' categories is a type of boundary making that Espeland and Sauder (2007) have termed 'commensuration'. Commensuration works by 'making irrelevant vast amounts of information, and by imposing on what remains the same form – a shared metric' (Espeland and Saunder, 2007: 17). Relevant to this last step of making knowledge irrelevant, Bowker and Star (1999) and Bonneuil et al. (2014) have illustrated relationships between sorting and categorization as social and institutional processes that transform lesser-status knowledge categories into 'invisible' knowledge. Showing the intermixing of *gen-ph* as both a category of relevance and of stigma makes explicit the connection between what Lamont and Molnár (2002) define as 'symbolic boundaries' and 'social boundaries' – the way that 'conceptual distinctions made by social actors to categorize objects, people, practices, and even time and space' is related to 'objectified forms of social differences manifested in unequal access to and unequal distribution of resources (material and nonmaterial) and social opportunities' (Lamont and Molnár, 2002: 168-169).

Given that moderators are now presented with the subject classifier results for all papers that pass through their hands, the issue of whether these results influence final decisions is of extreme importance, but lies outside the scope of the present work. Although moderators resisted the idea that the filters could strongly shape their decisions about submissions, it is well known that the outcomes of decision support systems introduce bias (Skitka et al., 1999, 2000). Moreover, the nature of the filters is to some extent self-actualising: updates to the filters are partly based on the outcomes of moderators' decisions, so the filters are implicated in a performativity self-loop.[20]

---

[19] The 'general physics' section was created by Ginsparg to capture papers by authors that had credentials, but which were uninteresting to any existing community. It replicates the infamous 'general physics' sessions at the American Physical Society's Annual March Meetings. These serve the purpose of capturing the work of amateur and independent physicist, most of which physicists would find uninteresting, as any and all members of the APS were entitled to present their work in some APS meeting. See for example https://www.reddit.com/r/Physics/comments/22cyeh/i_know_crackpot_submissions_happen_every_year_at/ (accessed 8 March, 2016).

[20] See MacKenzie (2003) and Barad (2014) for studies on performativity in relation to computer classification and sorting.

### ❧Arete: Gen-ph … at best❧

Currently, the two filters in conjunction are judged by Ginsparg to work well enough that they can identify non-classifiable submissions even better than can some moderators. Yet expulsion boundary work often also includes a process of stigmatization of what is left outside the boundary, in order to further legitimize control over the boundary. As a scientific institution, arXiv must describe its expulsion work in terms of 'scientific' motives. For example, a person unfamiliar with the culture of physics would have no reason to think that *gen-ph* is anything but one more subject category among many; nowhere on the arXiv site is *gen-ph* mentioned as a special or less important category. Yet, from the start, *gen-ph* was made to contain papers that have no specific audience and are therefore thought to be generally uninteresting to all specialized audiences. However, arXiv's expulsion boundary work is unmistakable:

> Ginsparg: We've actually had submissions to arXiv that are not spotted by the moderators but are spotted by the automated programme …. I put up this graph and have the two dimensional geography. All I was trying to do is build a simple text classifier and inadvertently I built what I call The Holy Grail of Crackpot Filtering.

'Crackpots' is a category with extreme stigma within professional physics. A 'crackpot' is a social recluse who is isolated from the 'intellectual' world of physics, an incompetent scientist, a fraud – a person whose chosen topic is so irrelevant that it is, as Wolfgang Pauli famously described, 'not even wrong' (see Baez, 1998; Gardener, 1957; Langmuir, 1989; Siegel, 2011; 't Hooft, 2003; see also Collins et al, manuscript, for a sociological analysis of these texts). The oral culture of physics, illustrated in online social networking sites, forums and blogs, evidences this stigma even further:

> But gen-ph is supposed to be for papers that no serious physicist would read.[21]

> [T]he Arxiv's approach to nonstandard papers is tagging instead of closing; what moderators did not wish to evaluate further or classify they tagged as [math.GM] and let the readers decide what, if anything, to make of those works.[22]

> There's also (usually in the general 'physics' category) a scattering of (crackpot) nontraditional science papers.[23]

---

[21] Comment by 'Kea' on Peter Woit's popular blog *Not Even Wrong*. 'Nielsen-Ninomiya and the arXiv' Available at: http://www.math.columbia.edu/~woit/wordpress/?p=2384 (accessed 8 March, 2016).

[22] Comment by 'T..', 'Is this question appropriate? Flawed proof of Green-Tao' thread, *Mathematics-meta Stack Exchange*. Available at: http://meta.math.stackexchange.com/questions/1433/is-this-question-appropriate-flawed-proof-of-green-tao (accessed 8 March, 2016).

[23] 'See what's really happening in physics – arXiv.org'. *Instructables* website. Available at: http://www.instructables.com/community/See-Whats-Really-Happening-in-Physics-arXivor/ (accessed 8 March, 2016).

> Blog publishing or gm-general mathematics publishing at ArXiv will threatens [sic] to mark him as a crackpot mathematician.[24]
>
> Yes, the permissive nature of arXiv means that some crackpots are submitting crazy theories (although it's surprisingly rare and the moderators push some in General Mathematics/General Physics).[25]
>
> I thought General Physics was the place destined to what arxiv mods considered as trash bin.[26]
>
> The preprint was re-classified from the professional hep-th archive to gen-ph, general physics, an archive mostly dedicated to laymen's fantasies.[27]
>
> Once you have an endorsement you still face the arXiv moderators who can turf your work out or move it to a category such as 'general physics' that nobody looks at.[28]

Though 'unofficial', the characterization of *gen-ph* as a crackpot category indicates that arXiv's boundaries not only put submissions into categories of interestingness and relevance, as could be claimed by arXiv, but also have serious consequences for submitters' reputations within wider physics audiences. Apart from being considered wholly uninteresting to the professional physicist, *gen-ph* papers are described in derogatory terms – 'fantastical', 'amateurish' and 'trash' – by the arXiv user culture. Ginsparg's use of the term 'crackpot' indicates that this situation is not unknown to arXiv. *gen-ph* therefore functions as an institutionalized compartment that can render papers invisible (Bonneuil et al. 2014; Collins, 2014), and it is 'the individual [who is] to be corrected' (Foucault, 1975). Following Bowker and Star, this analysis of arXiv's expulsion boundary work shows that, in the development of large-scale filtering and sorting systems, 'politically and socially charged agendas are often first presented as purely technical', and 'this leads to a naturalization of the political category' (Bowker and Star, 1999: 196).

**✥Sisyphean tales: Continuous reclassification into gen-ph✥**

The STS literature is rich with examples of fringe and oppositional science attempting to affect the mainstream: Bonneuil et al. (2014), Delborne (2008), Klein and Kleinman (2002), Martin (1988) and Wazeck (2013, 2014) discuss examples of the marginalization of minority and 'contrarian' views within mainstream science, while Ross (2010) discusses how the relationship between users and information systems is related to the reproduction of existing hierarchical structures. Similarly, in the final section of this paper I will consider the arXiv outsider's perspective, by illustrating consequences of being continuously stopped by the software filter, and then subjected to moderator reclassification and rejection.[29]

---

[24] Anonymous user, *Quora* thread, 'Suppose a passionate mathematical hobbyist thinks he has solved the Riemann Hypothesis…'. Available at https://www.quora.com/

[25] J. Desjardins-Proulx. The case for arXiv and a broader conception of peer-reviews. Available at: http://inngue.net/?q=node/330 (accessed 8 March, 2016).

[26] Comment by user 'Daniel de França', 'Levitating gas pipelines' post. *The physics arXiv* blog: http://arxivblog.com/?p=743 (accessed 8 March, 2016).

[27] E8 theory. Grenouille Bouille blog. Available at: https://grenouillebouillie.wordpress.com/2007/11/15/e8-theory/ (accessed 8 March, 2016).

[28] 'Crackpots'. Dave Goldberg's *A User's Guide to the Universe* blog. Available at: http://usersguidetotheuniverse.com/index.php/2010/06/18/crackpots/ (accessed 8 March, 2016).

In a controversial case in which a submission with unorthodox ideas is reclassified as *gen-ph*, the multiple interpretations of the category can have definite detrimental effects on how the author is viewed by the wider physics population, irrespective of whether the driving factor of arXiv's internal logic is content or relevance. This situation can have serious consequences for scientists who lean towards exploring unconventional topics or unorthodox methods, particularly when they are not in a secure academic position or work independently. To reiterate a point made earlier, here I do not refer to far outliers who take a primarily oppositional stance, but to individuals close to the blurry boundaries of legitimacy. Regarding a specific controversial reclassification into *gen-ph* of a paper that used atypical mathematical techniques (albeit from a researcher at a well-accredited institution), Ginsparg commented:

> Ginsparg: [Being reclassified into *gen-ph*] is not a devastating black mark but, you know, it looks kind of odd that somebody thought that this was not of interest to researchers in the research community …. You know, if it were me I don't know. I mean, I'd have to look at it, but if it were me making the [moderation] decision I would have a great deal of difficulty moving something from [someone from that institution], because they've already passed through so many levels of filtering.

I asked a scientist, who is familiar with and a proponent of the techniques used, and who has an interest in cranks, crackpots and unorthodox science, to comment on this reclassification. Upon reading the paper, he remarked, 'I don't conclude that [the author] is a "crank".' However, he also pointed out that the style of the paper could make readers suspicious:

> Scientist: Diagrams like the ones [used in the paper] don't really mean anything specific. Figure 1 first seems slightly helpful to me, but Figure 3 is the sort of thing that makes physicists think 'crackpot!' What does that slanted line mean, and what are those little arrows coming off it? There's no way to be sure. It seems to be his private 'visualization' of the text below. Physicists all have such mental images, but they know to keep them private or else explain them very precisely (e.g. Feynman diagrams) …. [The paper] is not speaking the language of modern mathematical physics at the level of sophistication that [similar physicists] are.[30]

Another excerpt from correspondence with this scientist reinforces the point that it was not only the content but also the style of a paper that could make the paper unappealing to mainstream readers:

> Scientist: There are different ways of talking about [this topic], some more modern than others, and the more modern ways tend to deliver more insight to those who understand them, while the older ways tend to involve a lot of matrices and explicit calculations, where you see that something is working but don't have the same high-level view as to why. … All this is of course a separate question from whether [these] ideas on physics are right – that is, relevant to our universe! I'm not trying to say that [mainstream

---

[29] The outlook follows a sociology of 'conflict' (Hård, 1993) in which technology is seen as 'a means for groups to retain or rearrange social relations', going beyond the usual constructivist view that considers the closure of controversies as the simple outcome of inter-group politics.

[30] Personal communication, 12 January 2015.

> techniques] are right while [this paper is] wrong …. Most mathematicians would be unable to follow [this paper] without a lot of work. And since [the authors] sometimes doesn't present their results as official 'theorems' – although there are theorems there – it's probably quite hard for most mathematicians to even see that there's something worth following …. They actually talk more like ordinary physicists (in their vocabulary), except for one huge difference: ordinary physicists don't know about this method.[31]

Sisyphus, an author referred to in the paper described above, recounted his own story of facing reclassification into *gen-ph,* and discussed the generalized academic stigma around similar atypical mathematical techniques:

> Sisyphus: In the 1970s, I began applying [these mathematical techniques] to theoretical physics …. For around 35 years I did this alone, primarily because I was addicted to the promise of the mathematics, but secondarily because those aware of my work considered such outré investigations professional suicide. They were not wrong. I was advised many times as a graduate student and post-doc to change my ways before it was too late.[32]

Although he had a track record of publications in the prestigious *hep-th* (theoretical HEP), his recent efforts to publish papers in arXiv resulted in rejections:

> Sisyphus: Until recently I've never had a problem getting anything posted to *hep-th*. However, a couple years ago I realized that acceptance of my core ideas could be used to solve one of the … biggest mysteries … in theoretical physics …. I wrote up a paper, the style of which I confess was glib and 'unprofessional'…. I eventually wrote a longer version, more detailed, and glib-free. This was submitted to *hep-th*, sent to be refereed, and outright rejected. Astonishingly the reason given was that I'd been writing papers of this sort (i.e., exploiting unsanctioned maths) for many decades, and the referee didn't see any reason *hep-th* should keep posting them.

That Sisyphus meant for the paper to be published specifically in the *hep-th* category is highly relevant, given that arXiv is the primary communication channel in HEP. In line with its arbitration policies, arXiv's administration decided that Sisyphus' paper would be posted if he managed to get it published in a mainstream journal. Lacking academic resources at that point, the decision motivated Sisyphus to give up.

Although Sisyphus has held temporary positions at around a dozen highly ranked universities, he has never secured a permanent position. He has occasionally written papers and has attended the odd mathematical conference. At one of these, it was 'strongly recommended' that the paper be published in the conference proceedings, and this fact enabled him to once more request that arXiv post the paper. However:

> Sisyphus: Once done – actually in print – I sent it off the arXiv, intending it to be listed in *hep-th*. I was informed that it would be listed, but in *gen-ph*. I assumed some gatekeeper, frustrated at finding such material could not be denied listing, made an effort to diminish its impact. I don't know …. It was NOT my intention that it be listed as *gen-ph* …. 4 or 5 years ago at a conference at Notre Dame in Indiana I was surprised to be introduced before my talk as 'the maverick'. My wife caught several knowing glances in

---

[31] Personal communication, 12 January 2015.

[32] Personal communication, 3 December 2014.

> the audience. I was not 'one of them'; I was something other. In retrospect I realized that yes, I was something of a maverick, but I had always thought it a choice. I now realize it was not a choice, any more than autism or dyslexia is a choice: I was just born that way.

Sisyphus' recollections illustrate the psychological and emotional effects of being denied posting rights. Despite holding academic positions at prestigious institutions, working on unorthodox topics and facing the rejection of the physics community eventually forced the author to a life outside academia.

> Sisyphus: That's the other side of the outsider maverick coin: when young, resistance to your ideas rankles almost constantly; and when you get older, because you've spent a life as an outsider, those on the inside are either ignorant of your work, or feel that they have a right – even duty – to ignore it anyway. There is no winning side to this coin …. That's why I am trying not to think about it much anymore. Trying.

Sisyphus's case, along with other cases of ex-physicists working outside academia I have analysed, illustrates that having a paper reclassified into *gen-ph* can have serious consequences for an author. By conflating interestingness and quality into a single category that at the same time renders papers 'invisible', non-trivial issues arise regarding the stigmatizing performativity of *gen-ph*.

## ❧Conclusions: Visibility and arXiv's role in the ecology of physics❧

Controversial reclassification cases are not a common occurrence in the day-to-day activity of arXiv. Moreover, even the foregoing authors whose papers were reclassified, along with the other arXiv outsiders I consulted, such as viXra founder Phils Gibbs (himself a sporadic arXiv user), admitted that the arXiv serves its purpose well most of the time. Yet despite its being initially conceptualized and designed as a pure dissemination system, some of arXiv's filtering can have negative implications for scientists whose papers are reclassified into *gen-ph*. arXiv thus functions as a certification site, by dividing its population into insiders and outsiders in the eyes of the wider physics community. The accounts from 'outsiders' to established research communities have evidenced how the success of being included can be highly dependent on access to institutional resources and on strong adherence to mainstream research programmes. arXiv's barriers are high enough that an author's lack of contact with a research community's form of life, or their choosing to work on unorthodox topics slightly outside mainstream interests, may cause filters to flag and moderators to reject or reclassify their submissions. The resulting perceived lack of diversity, particularly in HEP-related areas, has been the subject of some highly visible debates within the physics profession (e.g. Smolin, 2006).

As it has become increasingly important for communities like HEP, arXiv has transformed into a publishing outlet that is not just a preprint server, but also a primary community-wide communication space, particularly for newer generations.

> Moderator 3: What I have noticed is that, certainly, the younger people do use [arXiv] differently. I always try to get the journal version. For my students this is not true …. Many people just click on the arXiv link. In that sense it's really more and more like a tool of convenience.

> Moderator 4: arXiv means access for all, but you don't get a refereed version. Most of the publications have no issues. I know many students that don't even go to the official publication. I always do but I know many students that don't do it, they simply look at it, probably because they read it in a different way.

Exclusion from arXiv therefore becomes equivalent to exclusion from a research community's *de facto* main output conduit. Younger arXiv users and researchers who were interviewed for the project expressed the view that arXiv is increasingly being considered a certification channel, as opposed to just a rapid communications channel. Having a paper reclassified necessarily affects a researcher's visibility in a research culture in which online resources are often the preferred source of new knowledge. None of the moderators or journal editors interviewed could envision a case of a paper being rejected by a journal editor simply because it had been reclassified into *gen-ph*. Nonetheless, some of the younger interviewees feared this kind of case, precisely because *gen-ph* is generally acknowledged to be a category of social exclusion in physics culture.

In an online world where relevance and information flooding can only be handled by clever automated filtering, the matter of how these filters are constructed and enacted is of the utmost importance.[33] Although lacking the filter of peer-review to spot explicit technical adequacy, the sophisticated combination of software filtering, moderator scrutiny and peer-exposure in arXiv means that even individuals involved but not immersed in esoteric research cultures face significant barriers for posting on arXiv. During the course of this research, one case was encountered of a researcher who, like Sisyphus, was denied posting rights to the *hep-th* category until the paper in question could be published in a high profile journal. When the paper was accepted by a journal, it was posted to *hep-th*. The fact that in similar cases arXiv has chosen to use publication in a reputable mainstream journal as the *ultimately proxy* for quality control in the most controversial reclassification cases means that the bar for publishing in arXiv can at times be *higher* than for a mainstream journal. Yet important cases of top-level policymaking, misinformed about quality control in repositories, continue to appear. The influential report by Finch (2012), for example, erroneously stated that 'ArXiv is a preprint repository, for papers before they are submitted to a journal for peer review and publication' and that '[t]here is minimal filtering of incoming papers for quality control purposes' (Finch, 2012: 420).

arXiv's capacity to process vast amounts of submissions through automation and minimal human intervention has undoubtedly had a very positive effect on the dissemination of research, particularly in relation to communities that lack resources, like those in developing countries and independent researchers.[34] As academia moves toward different models, cultures and practices of open access, sociological analyses of arXiv boundary policing can not only address issues of balancing efficiency, expediency, fairness, openness and inclusion, but also exemplify how these issues cannot be separated from the internal social dynamics and politics of professional fields.

---

[33] The issue becomes more pressing for 'public' information retrieval systems like the increasingly popular Google Scholar, whose inner workings are, as Jacsó (2005) notes, completely obscure to users. See also De Winter et al. (2014), Schroeder (2014).

[34] See P. Ginsparg. 'The arXiv at 20'. Cornell University Library. Available at: http://www.cornell.edu/video/arxiv-at-20-the-price-for-success (accessed 24 June 2015).